\documentclass{PoS}

\title{$CP$ violation in charm decays at Belle}

\ShortTitle{$CP$ violation in charm decays at Belle}

\author{\speaker{B.~R.~Ko}\thanks{on behalf of the Belle Collaboration.}\\
        Korea University, Seoul, Republic of Korea\\
        E-mail: \email{brko@hep.korea.ac.kr}}
\abstract{We have searched for $CP$ violation of charmed mesons in the decays
  $D\rightarrow K^0_S P$, where $D$ denotes $D^0$ and $D^+_{(s)}$, and $P$
  denotes the pseudo-scalar mesons $\pi^+$, $K^+$, $\pi^0$, $\eta$, and
  $\eta'$. No evidence of $CP$ violation in these decays is observed. We also
  have measured the $CP$ asymmetry difference between the Cabibbo suppressed
  decay $D^+\rightarrow\phi\pi^+$ and the Cabibbo favored
  $D^+_s\rightarrow\phi\pi^+$ decays in the region of
  $|M(K^+K^-)-M^{\phi}_{PDG}|<$16 MeV/$c^2$. The measured asymmetry is
  corrected for the residual asymmetry due to detector effects, and the
  contributions of both $CP$ and forward-backward asymmetries are
  determined. These results are obtained on a large data sample collected at
  and near the $\Upsilon(4S)$ resonance with the Belle detector operating at
  the KEKB asymmetric-energy $e^+e^-$ collider.}

\FullConference{35th International Conference of High Energy Physics - ICHEP2010,\\
		July 22-28, 2010\\
		Paris France}

\begin{document}

Violation of the combined Charge-conjugation and Parity symmetries ($CP$) in
the standard model (SM) is produced by a non-vanishing phase in the
Cabibbo-Kobayashi-Maskawa flavor-mixing matrix~\cite{CKM}, where the violation
may be observed as a non-zero $CP$ asymmetry defined as 
\begin{equation}
  A^{D\rightarrow f}_{CP}=\frac
  {\Gamma(D\rightarrow f)-\Gamma(\bar{D}\rightarrow\bar{f})}
  {\Gamma(D\rightarrow f)+\Gamma(\bar{D}\rightarrow\bar{f})}
  \label{EQ:ACP}
\end{equation}
where $\Gamma$ is the partial decay width, $D$ denotes a charmed meson, and $f$
is a final state.

In the SM, the
charged charmed meson decays for which a significant non-vanishing $CP$
violation ($\mathcal{O}(0.1)$\% or lower~\cite{SMCP}) is expected are singly
Cabibbo-suppressed (SCS) decays in which there is both interference between two
different decay amplitudes and a strong phase shift from final state
interactions. The expected SM $CP$ violation in non-leptonic decay of the
neutral charmed meson is generated from interference of decays with and without
mixing in the absence of direct $CP$ violation in Cabibbo favored (CF) and
doubly Cabibbo suppressed (DCS) decays. The SM also predicts a $CP$ asymmetry
in the final states containing a neutral kaon that is produced via
$K^0-\bar{K}^0$ mixing even if no $CP$ violating phase exists in the charm
decays itself and we refer to it as $A^{K^0_S}_{CP}$. The magnitude of
$A^{K^0_S}_{CP}$ is $(0.332\pm0.006)$\%~\cite{PDG2010} if DCS decay
contributions are ignored. 

In this presentation, we report $CP$ asymmetries of charmed mesons in the decays
$D\rightarrow K^0_S P$, where $D$ denotes $D^0$ and $D^+_{(s)}$, and $P$
denotes the pseudo-scalar mesons $\pi^+$, $K^+$, $\pi^0$, $\eta$, and
$\eta'$~\cite{CC}. We also report the $CP$ asymmetry difference between SCS
decay $D^+\rightarrow\phi\pi^+$ and CF decay $D^+_s\rightarrow\phi\pi^+$ in the region of
  $|M(K^+K^-)-M^{\phi}_{PDG}|<$16 MeV/$c^2$. Among the decays listed above,
$D^+\rightarrow K^0_S K^+$ and $D^+_s\rightarrow K^0_S\pi^+$ are SCS decays and
others are mixtures of CF and DCS decays, where SM $CP$ violations described
above are expected. Interference between CF and DCS could generate
$\mathcal{O}(1)$\% of direct $CP$ asymmetry if unknown new physics processes
are responsible for additional weak phases~\cite{BIGI}. Physics beyond the SM
could also induce direct $CP$ asymmetry ($\mathcal{O}(1)$\%) in $D$ meson
decays~\cite{YAY}. Since $CP$ asymmetries expected by the SM in the decays
considered in this presentation is much smaller than $A^{K^0_S}_{CP}$,
observing $A_{CP}$ inconsistent with $A^{K^0_S}_{CP}$ would represent strong
evidence for processes involving physics beyond the
SM~\cite{BIGI}\cite{YAY}. The data were recorded at or near the $\Upsilon(4S)$
resonance with the Belle detector~\cite{BELLE} at the $e^+e^-$
asymmetric-energy collider KEKB~\cite{KEKB}. The sample corresponds to an
integrated luminosity of 673/791/854 fb$^{-1}$ depending on the decay mode.

\begin{figure}[htbp]
\begin{center}
  \includegraphics[width=0.6\textwidth]{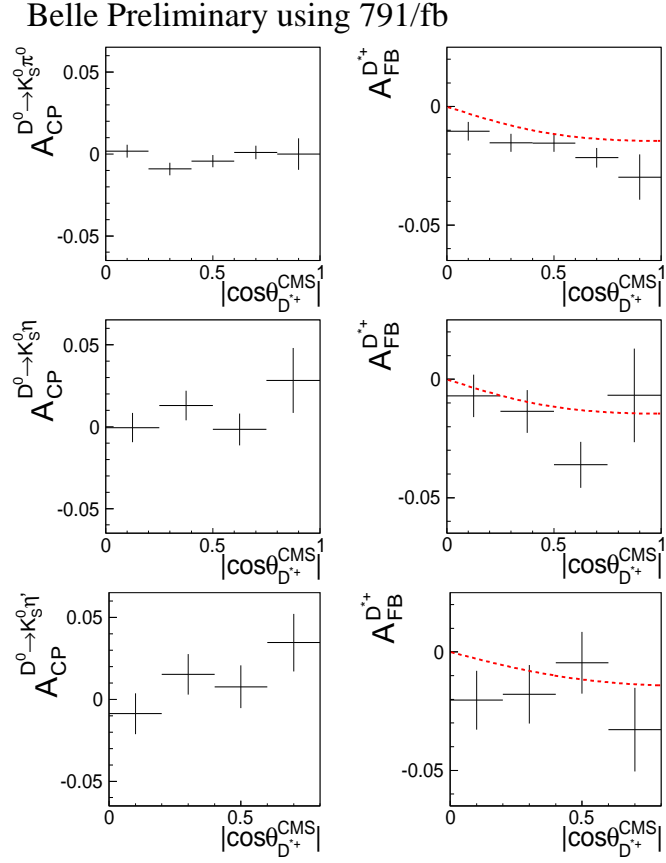}
\caption{Preliminary results of $A_{CP}$ (left) and $A_{FB}$ (right) values as
  a function of $|\cos\theta^{\rm CMS}_{D^{*+}}|$. Top plots are for
  $K^0_S\pi^0$, middle plots for $K^0_S\eta$, and bottom plots for $K^0_S\eta'$
  final states. The dashed curves show the leading-order prediction for
  $A^{c\bar{c}}_{FB}$.}
\label{FIG:ACPD0}
\end{center}
\end{figure}

We determine the quantity $A^{D\rightarrow f}_{CP}$ defined in
Eq.~(\ref{EQ:ACP}) by measuring the asymmetry in the signal yield
\begin{equation}
  A^{D\rightarrow f}_{\rm rec}~=~\frac
  {N_{\rm rec}^{D\rightarrow f}-N_{\rm rec}^{\bar{D}\rightarrow\bar{f}}}
  {N_{\rm rec}^{D\rightarrow f}+N_{\rm rec}^{\bar{D}\rightarrow\bar{f}}}    
  ~=~A^{D\rightarrow f}_{CP}~+~A_{\rm other},
  \label{EQ:ARECON}
\end{equation}
where $N_{\rm rec}$ is the number of reconstructed decays. $A_{\rm other}$ is
asymmetry other than $A_{CP}$ and it contains the forward-backward asymmetry
($A_{FB}$) due to $\gamma^{*}-Z^0$ interference in $e^+e^-\rightarrow c\bar{c}$
and the other is a detection efficiency asymmetry between positively and
negatively charged hadrons and the latter depends on decay mode. With
assumption the $A_{FB}$ is the same for all charmed mesons, we correct for
$A_{\rm other}$ using a large statistics of real data samples. The detailed
correction procedures are described in
Refs.~\cite{ACPKSH}\cite{D0hhBelle}\cite{D0hhBaBar}. Once we correct for
$A_{\rm other}$, then $A^{D\rightarrow f}_{CP}$ is obtained in bins of
corresponding phase spaces (shown in Fig.~\ref{FIG:ACPD0}) and the measured
$A_{CP}$ values are listed in Table~\ref{TABLE:SUMMARY}.

\begin{table}[htbp]
\begin{center}
\caption{Summary of the $A_{CP}$ measurements. The first uncertainties are
  statistical and the second are systematic. The $\ddagger$ is the total
  uncertainty. The $\dagger$'s are preliminary results.}
\label{TABLE:SUMMARY}
\begin{tabular}{cccc} \hline \hline
Decay Mode                    &$A_{CP}$ (\%) (Belle)&$A_{CP}$ (\%) (current
world best or world average)&$A^{K^0_S}_{CP}$ (\%) \\ \hline 
$D^+\rightarrow K^0_S\pi^+$   &$-0.71\pm0.19\pm0.20$&$-1.3\pm0.7\pm0.3$   &$-0.332$\\ 
$D^+\rightarrow K^0_S  K^+$   &$-0.16\pm0.58\pm0.25$&$-0.2\pm1.5\pm0.9$   &$-0.332$\\ 
$D^+_s\rightarrow K^0_S\pi^+$ &$+5.45\pm2.50\pm0.33$&$+16.3\pm7.3\pm0.3$  &$+0.332$\\ 
$D^+_s\rightarrow K^0_S K^+$  &$+0.12\pm0.36\pm0.22$&$+4.7\pm1.8\pm0.9$   &$-0.332$\\ \hline
$D^0\rightarrow K^0_S\pi^0$   &$-0.28\pm0.19\pm0.10$$^\dagger$&$+0.1\pm1.3$$^\ddagger$&$-0.332$\\ 
$D^0\rightarrow K^0_S\eta$    &$+0.54\pm0.51\pm0.16$$^\dagger$&N.A.                 &$-0.332$\\ 
$D^0\rightarrow K^0_S\eta'$   &$+0.90\pm0.67\pm0.14$$^\dagger$&N.A.                 &$-0.332$\\ \hline \hline 
\end{tabular}     
\end{center}
\end{table}

\begin{figure}[htbp]
\begin{center}
  \includegraphics[width=0.85\textwidth]{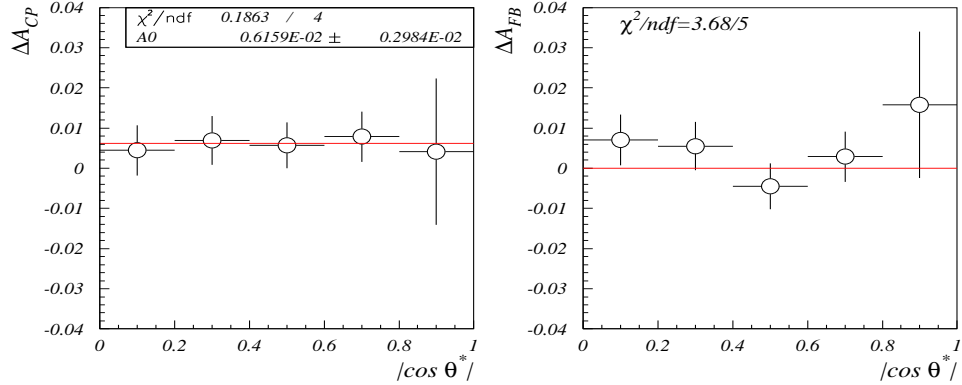}
  \caption{Preliminary results of $\Delta A_{CP}$ (left) and $\Delta A_{FB}$
  (right) values as a function of $|\cos\theta^{*}|$. The line in left plot
  shows the fit with a constant and that of right shows the hypothesis test for
  a null $\Delta A_{FB}$ hypothesis.}
  \label{FIG:DELACP}
\end{center}
\end{figure}

The $CP$ asymmetry difference between SCS decay $D^+\rightarrow\phi\pi^+$ and
CF decay $D^+_s\rightarrow\phi\pi^+$ ($\Delta A_{CP}$) is obtained by
subtracting $A^{D^+_s\rightarrow\phi\pi^+}_{\rm rec}$ from
$A^{D^+\rightarrow\phi\pi^+}_{\rm rec}$ since the kinematics of
$D^+\rightarrow\phi\pi^+$ and $D^+_s\rightarrow\phi\pi^+$ are quite similar
with each other. Besides the $\Delta A_{CP}$, the production difference between
$D^+$ and $D^+_s$ ($\Delta A_{FB}$) is also obtained by the
subtraction. Figure~\ref{FIG:DELACP} shows the measured $\Delta A_{CP}$ and
$\Delta A_{FB}$ in bins of corresponding phase space in the region of
$|M(K^+K^-)-M^{\phi}_{PDG}|<$16 MeV/$c^2$. By fitting the $\Delta A_{CP}$
points with a constant, we obtain a preliminary result of $\Delta
A_{CP}=(0.62\pm0.30\pm0.15)\%$ where the first uncertainty is statistical and
the second is systematic. The $\Delta A_{FB}$ plot in Fig.~\ref{FIG:DELACP}
shows no significant difference between forward-backward asymmetries in the
production of the $D^+$ and $D^+_s$ mesons.

In summary, we have searched for $CP$ violation in several charm decays. No
evidence for $CP$ violation is observed at sensitivities greater than 0.2\%
depending on the decay mode. We also find no significant difference between
forward-backward asymmetries in the production of the $D^+$ and $D^+_s$ mesons.


\begin{thebibliography}{99}
\bibitem{CC}
Throughout this Letter the charge conjugate decay mode is also implied unless
stated otherwise.
\end{thebibliography}
\end{document}